\documentclass[aps,prb,twocolumn,showpacs,floatfix,superscriptaddress,amsmath,amssymb]{revtex4-2}
\usepackage{amsfonts}
\usepackage{mathrsfs}
\usepackage{amsmath}
\usepackage{xcolor}
\usepackage{graphicx}
\usepackage{bm}
\usepackage{amssymb}
\usepackage{xspace}
\usepackage{epstopdf}
\usepackage{dcolumn}
\usepackage{longtable}
\usepackage{multirow}
\usepackage{float}
\usepackage{comment}
\usepackage[colorlinks=true, letterpaper=true, pdfstartview=FitV,  linkcolor=blue, citecolor=blue, urlcolor=blue]{hyperref}

\makeatother

\begin{document}

\title{Tunable linear and nonlinear anomalous Hall transport in two-dimensional CrPS$_{4}$ }

\author{Lulu Xiong}
\affiliation{Institute of Applied Physics and Materials Engineering, University of Macau, Taipa, Macau SAR, China}

\author{Jin Cao}
\email{caojin.phy@gmail.com}
\affiliation{Institute of Applied Physics and Materials Engineering, University of Macau, Taipa, Macau SAR, China}

\author{Fan Yang}
\affiliation{School of Physics, Beihang University, Beijing 100191, China}

\author{Xiaoxin Yang}
\affiliation{Institute of Applied Physics and Materials Engineering, University of Macau, Taipa, Macau SAR, China}

\author{Shen Lai}
\email{laishen@um.edu.mo}
\affiliation{Institute of Applied Physics and Materials Engineering, University of Macau, Taipa, Macau SAR, China}

\author{Xian-Lei Sheng}
\affiliation{School of Physics, Beihang University, Beijing 100191, China}
\affiliation{Peng Huanwu Collaborative Center for Research and Education, Beihang University, Beijing 100191, China}

\author{Cong Xiao}
\email{congxiao@fudan.edu.cn}
\affiliation{Interdisciplinary Center for Theoretical Physics and Information Sciences(ICTPIS), Fudan University, Shanghai 200433, China}

\author{Shengyuan A. Yang}
\affiliation{Research Laboratory for Quantum Materials, Department of Applied Physics, The Hong Kong Polytechnic University, Hong Kong SAR, China}

\begin{abstract}
Few-layer CrPS$_{4}$ is a two-dimensional (2D) magnetic material with excellent stability in ambient environment, which attracted significant interest in recent research. Here, via first-principles calculations, we show that 2D CrPS$_{4}$ hosts a variety of intriguing anomalous Hall transport phenomena, owing to its layer-dependent magnetism and symmetry character.
Monolayer CrPS$_{4}$ can display a sizable linear anomalous Hall effect, while its nonlinear anomalous Hall response is forbidden. In contrast, bilayer CrPS$_{4}$ can produce pronounced intrinsic nonlinear anomalous Hall response from Berry-connection polarizability, in the absence of linear anomalous Hall effect. We clarify that the large peaks in these responses originate from gapped Dirac points in the band structure. Furthermore, we show that bilayer CrPS$_{4}$ also hosts a layer Hall effect, where the charge Hall current in the two layers are equal in magnitude but flow in opposite directions. This effect can be detected as an anomalous Hall signal induced by an applied gate field. Finally, even-layer CrPS$_{4}$ also exhibits pronounced in-plane anomalous Hall effect, where the Hall signal is linearly proportional to the applied in-plane magnetic field. Our findings unveil the rich and interesting Hall transport phenomena in 2D CrPS$_{4}$ magnets, suggesting its potential in electronic and spintronic device applications.

\end{abstract}

\maketitle

\section{Introduction}
The various Hall effects are of fundamental importance in materials science and device applications. A prominent example is the (linear) anomalous Hall effect (AHE), where a Hall voltage is induced by a longitudinal current flow in a magnetic material~\cite{RevModPhys.82.1539}. It offers an important characterization of magnetic materials and a convenient method for detecting magnetization dynamics.
A key advance in the past two decades is the recognition that AHE contains an intrinsic contribution, i.e., an response determined solely by the material's band structure, which
manifests an intriguing band geometric property, namely, the Berry curvature of electronic bands~\cite{PhysRevLett.88.207208,onoda2002topological}. AHE is a time-reversal-odd ($\mathcal{T}$-odd) effect, meaning that reversal of magnetic moments in the system (corresponding to a time-reversal operation) must flip the sign of the response. The $\mathcal{T}$-odd Hall effect was also discovered in nonlinear response of magnets, giving a nonlinear Hall current quadratic in the driving $E$ field, i.e., $j_\text{H}\propto E^2$~\cite{PhysRevLett.112.166601,PhysRevLett.127.277201,PhysRevLett.127.277202}.
Importantly, this nonlinear AHE also contains an intrinsic contribution, which is determined by another band geometric quantity, the Berry-connection
polarizability (BCP)~\cite{PhysRevLett.127.277202,PhysRevB.105.045118}, and it can be further connected to the quantum metric of band structure~\cite{PhysRevLett.127.277201, Gao_2023, FENG2025100040}. Note that
because of its $\mathcal{T}$-odd nature,
this intrinsic nonlinear AHE can occur only in magnetic materials; it
is distinct from the $\mathcal{T}$-even nonlinear AHE that has been widely studied in nonmagnetic materials and does not have an intrinsic contribution~\cite{PhysRevLett.115.216806,ma_observation_2019,kang2019nonlinear,du_nonlinear_2021}.
So far, the intrinsic nonlinear AHE has been experimentally detected in only a few materials, such as few-layer MnBi$_{2}$Te$_{4}$~\cite{ Gao_2023, Wang_2023} and Mn$_{3}$Sn~\cite{han_room-temperature_2024}.
However, these systems each has certain drawbacks that are not ideal for studying intrinsic nonlinear AHE. For example, MnBi$_{2}$Te$_{4}$ preserves a threefold rotational symmetry along $z$ direction, which forbids intrinsic nonlinear AHE in the $xy$-plane~\cite{PhysRevLett.127.277201,PhysRevLett.127.277202},
and some extrinsic mechanism, e.g., by using low-symmetry substrate or capping layers to break the $C_{3z}$ symmetry~\cite{Gao_2023}, is need for observing the intrinsic nonlinear AHE. In such a case, the detected response does not represent an intrinsic property of the magnet itself, but should be attributed to the interface or the whole combined heterostructure. For Mn$_{3}$Sn, it also requires the strong interfacial effect at its interface with heavy-metal substrate~\cite{han_room-temperature_2024}. In addition,
Mn$_{3}$Sn is not a van der Waals (vdW) layered material, making it more demanding in experimental fabrication of desired interfacial configurations. Therefore, at current stage, it is still an urgent task to find suitable magnetic materials that host intrinsic nonlinear AHE as its bulk property (i.e.,
without requiring interfacial effects). And it would be even better if such materials belong to the vdW 2D materials family.

Indeed, 2D magnetic materials have been a focus of recent research~\cite{burch2018magnetism, mak2019probing,kurebayashi2022magnetism}. Due to the reduced dimensionality, a big advantage of 2D magnets is that their properties can be flexibly tuned, e.g., by gating, doping, strain, external fields, and forming vdW heterostructures. Combined with magnetism, 2D magnetic materials
are believed to hold tremendous potential for the next-generation information devices. Several material classes, like
CrI$_3$~\cite{huang2017layer}, Fe$_3$GeTe$_2$~\cite{deng2018gate,fei2018two}, Cr$_2$Ge$_2$Te$_6$~\cite{gong2017discovery}, and MnBi$_2$Te$_4$~\cite{PhysRevX.11.011003,ovchinnikov2021intertwined}, have been realized.
Nevertheless, it was noted that most of 2D magnets studied so far are not stable in ambient environment, posing difficulties for applications. In this regard, the recently realized 2D magnet CrPS$_{4}$ starts to attract increasing attention because of its excellent air stability~\cite{son2021airb}. It was shown that its magnetic ordering can persist down to monolayer limit~\cite{son2021airb}.
The fascinating properties of CrPS$_{4}$, such as spin dynamics, layer-dependent magnetism, optical response, electric and thermal transport properties, have been extensively investigated in recent works~\cite{son2021airb, pei2016spin, PhysRevB.94.195307, Lee2017Structural, kim_crossover_2019, peng2020magneticbbbb,calder2020magnetic,gu_photoluminescent_2020,Neal_2021,PhysRevMaterials.5.034005, wu2022magnetotransport, riesner2022temperature,kim2022photoluminescence,harms2022metal, wu2023gate, wu2023magnetism,huang2023layer, qi_giant_2023,PhysRevB.108.024405,cheng2024quantum,Liu_2024adv,PhysRevX.14.041065,Tunneling2024, yao_switching_2025,ho_imaging_2025}. As a semiconductor, the type and density of its charge carriers can be conveniently controlled by gating~\cite{wu2023gate,wu2023magnetism}.  Devices based on 2D CrPS$_{4}$, such as field effect transistors and magnetic tunnel junctions, have been studied in experiment~\cite{wu2023magnetism,huang2023layer, cheng2024quantum}. However, the anomalous Hall transport properties, which is an important characteristic of magnetic materials, have not been studied in
2D CrPS$_{4}$ yet.

In this work, we reveal that 2D CrPS$_{4}$ actually offers a good platform to study a variety of interesting Hall effects, especially the intrinsic nonlinear AHE.
First, 2D CrPS$_{4}$ with odd number of layers is ferromagnetic (FM), which allows linear AHE.
Second, even-layer CrPS$_{4}$ has a compensated antiferromagnetic (AFM) ordering, which has vanishing linear AHE
but allows intrinsic nonlinear AHE. As mentioned above, the symmetry condition of intrinsic nonlinear AHE is stringent: It requires
the absence of any out-of-plane rotation axis~\cite{PhysRevLett.127.277201,PhysRevLett.127.277202}. CrPS$_{4}$ stands out in this respect, as it naturally satisfies the condition due to its low-symmetry monoclinic crystal structure.
Based on first-principles calculations and taking monolayer and bilayer CrPS$_{4}$ as representatives, we quantitatively evaluate these intrinsic Hall response coefficients.
For monolayer CrPS$_{4}$, we show that it has a sizable anomalous Hall conductivity in the case of hole doping. Its value can be strongly tuned by the magnetization direction. For bilayer CrPS$_{4}$, we show that the intrinsic nonlinear AHE is
determined by a single response coefficient, which can achieve a large value $\sim 1 $ mA/V$^2$ under hole doping.
The origin of the large contribution in BCP is traced to a band anti-crossing region in valence bands.
Furthermore, we unveil that bilayer CrPS$_{4}$ actually hosts a layer Hall effect, where the linear charge Hall current in the two layers are flowing in opposite directions. This effect can be probed as a net linear charge Hall signal induced by
a gate electric field. The resulting current flow has a dominant distribution in one of the layers and the resulting Hall signal is odd in the applied gate field. Finally, we propose that a linear AHE signal can also be generated in even-layer
CrPS$_{4}$ by an in-plane magnetic field and the response is odd in the magnetic field, an effect known as the in-plane AHE.
Based on reported experimental data, we show that the result can be quite large under moderate applied field strength.
Our work reveals interesting Hall transport effects in the 2D air-stable magnet CrPS$_{4}$, which form the basis for  novel device applications in electronics and spintronics.

\section{Computation method}
Our first-principles calculations based on density functional theory (DFT) were performed by using Vienna \emph{ab-initio} simulation package~\cite{Kresse1993Ab,Kresse1996Efficiency,Kresse1996Efficient}, employing the projector augmented wave pseudo-potentials~\cite{Bloechl1994Projector}. The exchange-correlation energy was treated with the generalized-gradient approximation~\cite{Perdew1992Atoms} in the scheme of the Perdew-Burke-Ernzerhof realization~\cite{Perdew1996Generalized}. To account for vdW interactions, the DFT-D3 method was adopted~\cite{Grimme2010consistent}. The Dudarev \textit{et al.}\textquoteright s approach~\cite{Dudarev1998Electron} was used to treat possible correlation effects of Cr-$3d$ orbitals. Following Ref.~\cite{son2021airb}, we took effective $U$ value to be $2$~eV, which gives magnetic moment and bandgap values consistent with experiment.
The plane-wave cutoff energy was set to 400~eV, and a $k$-point mesh with size $12\times12\times1$ was used for the Brillouin zone (BZ)  sampling. For 2D layers, a vacuum space of thickness 18~$\text{\AA}$ was added to suppress artificial interactions
among periodic images.
The convergence criteria for the total energy and the force were set to $10^{-6}$~eV and 0.01~eV$/\text{\AA}$, respectively. The spin-orbit coupling was included in all calculations. Based on DFT results, \textit{ab-initio} tight-binding models were constructed by using the Wannier90 package~\cite{Mostofi2014updated}, which were then used for computing band geometric quantities and transport coefficients.

\begin{figure}[t]
  \includegraphics[width=8.5cm]{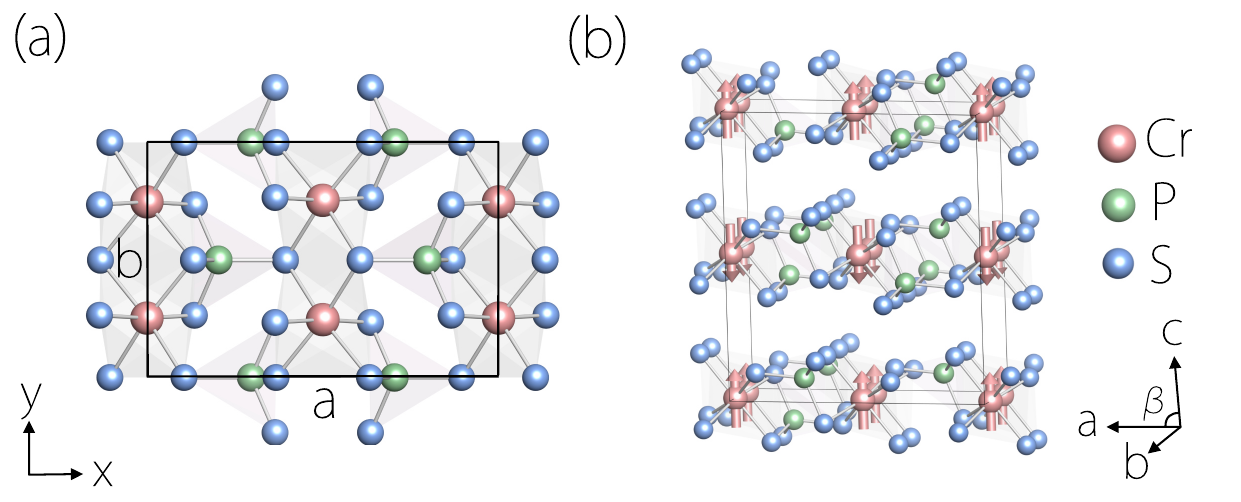}
  \caption{\label{fig1}(a) Top view of a CrPS$_4$ layer. The black rectangle marks the conventional cell.
  (b) Lattice structure of bulk CrPS$_4$. It has a $A$-type AFM ground state. The local moments are in the out-of-plane direction, as indicated by the red arrows in the figure. }
\end{figure}


\section{Structure and magnetic ordering}

Bulk CrPS$_{4}$ is a vdW layered material with monoclinic crystal structure (see Fig.~\ref{fig1}). It was first synthesized and studied in the 1970s~\cite{Diehl_a14649,Toffoli_a14371, Louisy1978Physical}. A recent experimental characterization reported that its lattice parameters (for the conventional cell) are $a=10.856\,\text{\AA}$, $b=7.247\,\text{\AA}$, and $c=6.135\,\text{\AA}$~\cite{peng2020magneticbbbb}. As shown in Fig.~\ref{fig1}(a), in a layer, each Cr cation is surrounded by six S anions, forming a CrS$_6$ octahedron with slight Jahn-Teller distortion. These CrS$_6$ octahedra form quasi-1D chains along $y$ direction in Fig.~\ref{fig1}(a), which are further inter-connected by PS$_4$ tetrahedra, resulting in a rectangle lattice in the 2D plane.
Due to the relatively weak vdW interaction, a slight slip along a axis direction occurs between neighboring layers, causing angle $\beta=88.116^{\circ}$, which deviates slightly from $90^{\circ}$ (see Fig.~\ref{fig1}(b))

It is noted that two initial experiments in 1977 reported slightly different crystal structures of bulk
CrPS$_{4}$ ~\cite{Diehl_a14649,Toffoli_a14371}. Using X-ray diffraction method, Ref.~\cite{Diehl_a14649} reported a crystal structure with $C2/m$ symmetry, i.e., with preserved inversion symmetry. On the other hand, Ref.~\cite{Toffoli_a14371} reported a very close structure but with weak inversion symmetry breaking, leading to a $C2$ symmetry. Comparing the two, the symmetry difference is associated with slight atomic position shift in each layer, mainly on the S site. Most subsequent research works on CrPS$_{4}$ just quote the result, $C2/m$~\cite{son2021airb,PhysRevMaterials.5.034005,yao_switching_2025,small_poto_2019,susilo_band_2020,PhysRevB.108.155133,HuLiLi2024,asada_nonlinear_2025} or $C2$~\cite{PhysRevB.94.195307,kim_crossover_2019,peng2020magneticbbbb,calder2020magnetic,gu_photoluminescent_2020,kim2022photoluminescence,qi_giant_2023,PhysRevB.108.024405,Liu_2024adv,ho_imaging_2025,Louisy1978Physical, kim_polarized_2021,Li_weiaaa_2023,nano13061128}, from either one of Refs.~\cite{Diehl_a14649,Toffoli_a14371}. And in fact, because of the two structures are very close, they do not make any qualitative difference
in explaining experimental results reported so far (within experimental accuracy).
Our DFT calculations indicate that at least for 3D bulk CrPS$_{4}$, monolayer CrPS$_{4}$, and bilayer CrPS$_{4}$, the $C2/m$ structure is energetically more stable than the $C2$ structure.
Our test calculation also shows that the small structural difference makes little effect on the Hall transport properties, e.g., the linear AHE coefficient between the two are almost identical.
Based on these considerations, we will take the $C2/m$ crystal structure for the investigation here.
It follows that the bulk structure of CrPS$_{4}$ preserves the spatial inversion symmetry $\mathcal{P}$ to a good extent. However, due to the in-plane anisotropy as described above, there is no vertical rotation axis (only a twofold axis along $y$ exists).

\begin{figure}
  \includegraphics[width=8.6cm]{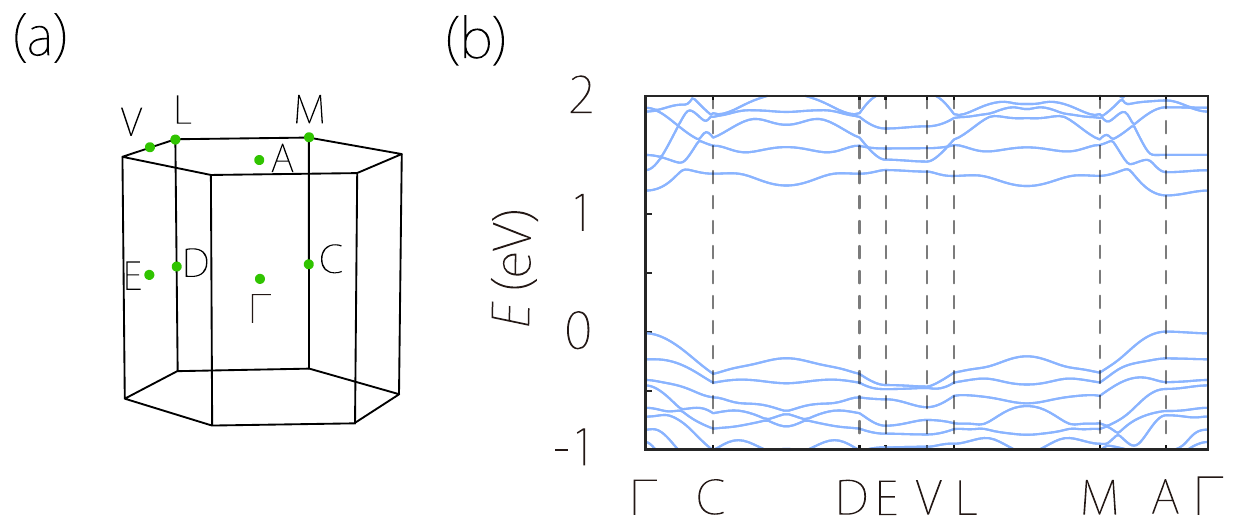}
  \caption{\label{fig2}(a) Brillouin zone and (b) calculated band structure for bulk CrPS$_4$.}
\end{figure}

The magnetic ground state of CrPS$_{4}$ has been well established in previous studies~\cite{son2021airb,peng2020magneticbbbb}. The magnetism comes mainly from the Cr$^{3+}$ ions with a magnetic moment about $3.0\mu_B$. The magnetic coupling within each layer is FM and the moments are in the out-of-plane direction; whereas the coupling between neighboring layers is AFM. This results in an overall $A$-type AFM ordering for bulk CrPS$_{4}$, with N\'{e}el vector along $z$ axis, as shown in Fig.~\ref{fig1}(b). The N\'{e}el temperature reported in experiment is about 35 to 38 K~\cite{son2021airb, Lee2017Structural,peng2020magneticbbbb,calder2020magnetic,huang2023layer}.

As a vdW layered material, 2D few layers can be easily exfoliated from bulk CrPS$_{4}$. Experimentally,
2D CrPS$_{4}$ with thickness down to monolayer has been successfully fabricated, via mechanical exfoliation or thermal atomic layer etching method~\cite{son2021airb}. Compared to other 2D magnets, a big advantage of this material is its good stability in air. It was shown that the magnetism is robust down to monolayer limit, and the ground state exhibits a layer-dependent alternation between two configurations. The odd layers are uncompensated antiferromagnet with a net magnetization along $z$ (FM for monolayer), whereas the even layers feature a fully compensated AFM state. The magnetic transition temperature shows a slight decrease when thickness decreases. The transition temperature for monolayer is found to be $\sim22$~K~\cite{son2021airb}. Taking into account the ground state magnetic configuration, the odd layers have $2^{\prime}/m^{\prime}$ magnetic point group, whereas the even layers belong to $2/m^{\prime}$. Their key difference is that
the odd layers preserve $\mathcal{P}$ (not $\mathcal{PT}$), but the even layers preserve $\mathcal{PT}$ (not $\mathcal{P}$).
This has important consequences on the type of Hall effects allowed in the system, as we shall discuss in a while.

The band structure for bulk CrPS$_{4}$ in the AFM ground state obtained from our DFT calculation is plotted in Fig.~\ref{fig2}(b).
The result shows that CrPS$_{4}$ is a magnetic semiconductor. The obtained bandgap value $\sim$1.21~eV agrees very well with the recent experimental result ($\sim$1.31~eV)~\cite{Lee2017Structural}. Our calculated magnetic moment $\sim 3.0\mu_{B}$ on Cr site also agrees with several previous experiments~\cite{son2021airb,peng2020magneticbbbb}. These support the validity of our computation method. In the case of few layers, we find that they are also magnetic semiconductors. The local magnetic moments remain nearly unchanged compared to the bulk. These features are consistent with previous studies.
In the following, we shall focus on  monolayer and bilayer CrPS$_{4}$ to study their various Hall responses.

\section{Anomalous Hall response in monolayer}

\begin{figure}[t!]
\includegraphics[width=8.6cm]{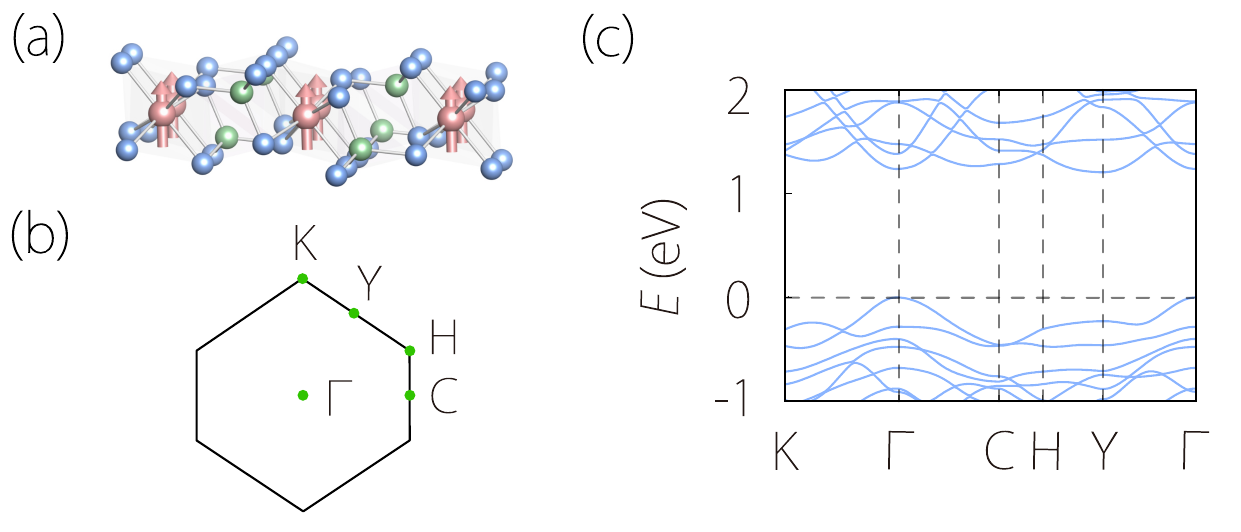}
\caption{\label{fig3}(a) Lattice structure of monolayer CrPS$_4$, which has FM ordering with out-of-plane magnetization. (b) Brillouin zone of the monolayer. (c) Band structure for monolayer CrPS$_4$ in the FM ground state.}
\end{figure}

Due to the $A$-type magnetic ordering, odd-layer CrPS$_{4}$ has an uncompensated net magnetization. For monolayer CrPS$_{4}$, this reduces to the FM ordering, with all magnetic moments in the out-of-plane direction ($z$ direction in our setup).
The band structure for this ground state is plotted in Fig.~\ref{fig3}(c) The result shows a semiconductor with a bandgap $\sim$1.20~eV, which agrees with previous studies~\cite{son2021airb}.

As we noted above, the magnetic point group for monolayer CrPS$_{4}$ (with FM ordering) is
$2^{\prime}/m^{\prime}$, which consists of two elements: $\mathcal{P}$ and $\mathcal{T}\mathcal{M}_{y}$ ($\mathcal{M}_{y}$ is the mirror normal to $y$). This symmetry allows for a nonzero
$\sigma_{xy}$ for linear AHE. On the other hand, any nonlinear charge transport, including nonlinear Hall effect, is forbidden.

Here, we compute the intrinsic AHE response. The corresponding anomalous Hall conductivity is given by~\citep{PhysRevLett.88.207208,PhysRevLett.92.037204}
\begin{equation}\label{AHE}
\sigma_{xy}=\frac{e^2}{\hbar}\int[d\boldsymbol{k}]f_{n\bm k}\Omega_{z}^{n}(\boldsymbol{k}),
\end{equation}
where $[d\boldsymbol{k}]$ is the short notation of $\sum_{n} d\boldsymbol{k}/(2\pi)^{2}$, $n$ is the band index, $f_{n\bm k}$ is the Fermi-Dirac distribution function, and
\begin{equation}\label{BC}
  \Omega_{z}^{n}(\boldsymbol{k})=-2\hbar^2 \text{Im}\sum_{m\neq n} \frac{(v_x)_{nm}(v_y)_{mn}}{(\varepsilon_n-\varepsilon_m)^2}
\end{equation}
is the Berry curvature of Bloch band structure, with $\varepsilon_n$ being the band energy and $(v_i)_{mn}$ being the interband velocity matrix element. For simple notations, we suppress the $k$ dependence in Eq.~(\ref{BC}) above. From Eq.~(\ref{AHE}), one can see that this AHE response is intimately connected to Berry curvature of occupied states.

\begin{figure}[t!]
  \includegraphics[width=8.6cm]{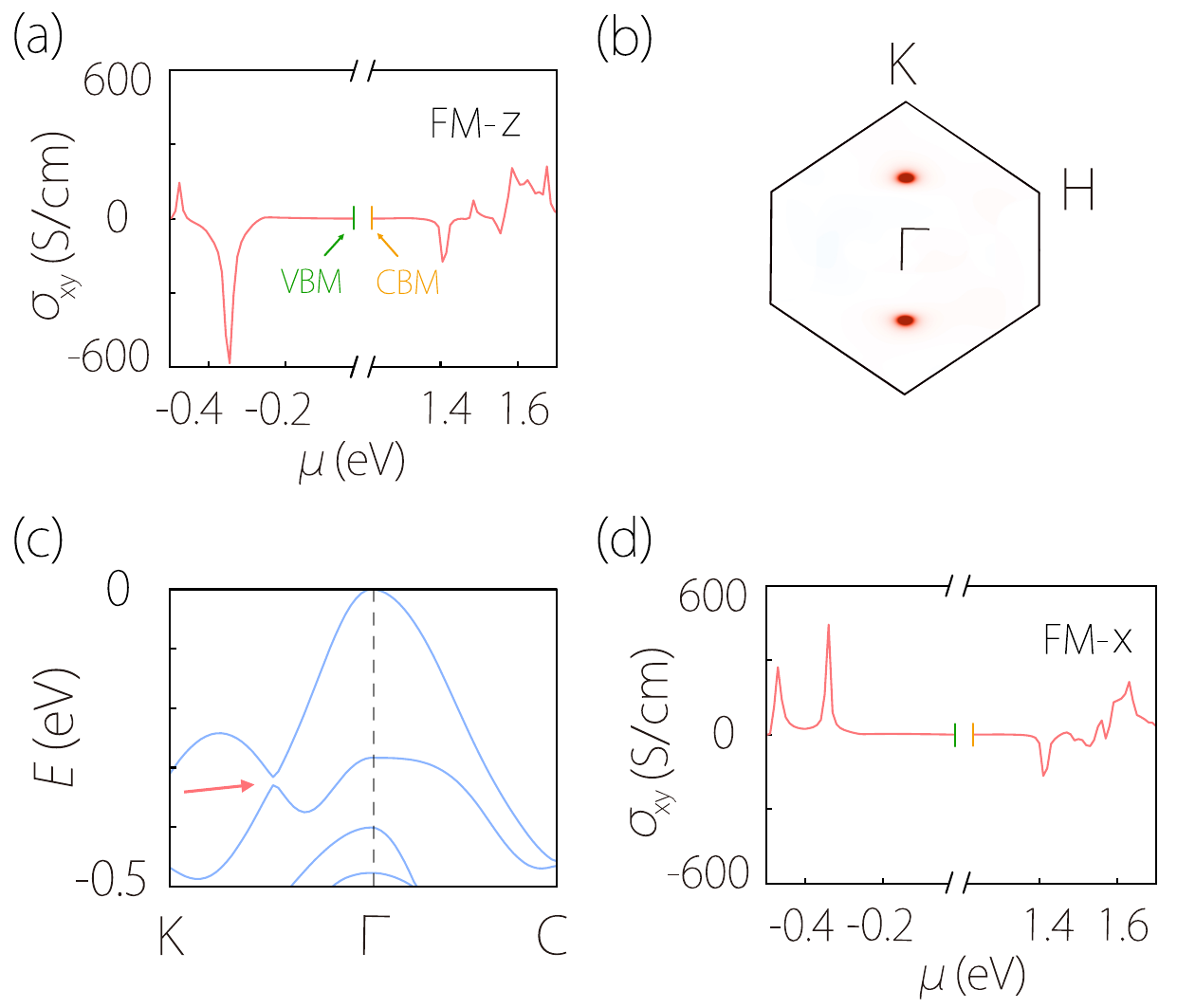}
  \caption{\label{fig4} (a) Calculated intrinsic anomalous Hall conductivity for monolayer CrPS$_4$ as a function of chemical potential. (b) Distribution of Berry curvature for states below $\mu=-0.32$~eV. (c) The hot spots in (b) correspond to the gapped Dirac points in band structure. The red arrow indicates one point. The other one is related to this one by inversion symmetry. (d) The result of intrinsic anomalous Hall conductivity when the magnetization is rotated to be along the $x$ direction.}
\end{figure}

In Fig.~\ref{fig4}(a), we show the calculation result of $\sigma_{xy}$ as a function of the chemical potential $\mu$. One can observe several peaks in both hole-doped and electron-doped regions. The peak at $\mu\sim -0.32$~eV can reach up to $\sim -600$~S/cm, which is a quite large value. To understand the origin of this peak, we calculate the Berry curvature distribution in $k$ space summed over states below $-0.32$~eV. The result is shown in Fig.~\ref{fig4}(b). One can see that large Berry curvature comes from two hot spots sitting on the $\Gamma$-$K$ path. These hot spots correspond to the small-gap regions indicated in Fig.~\ref{fig4}(c). It should be noted that here, the two bands do not cross, instead, they form an anti-crossing with a small gap $\sim0.014$~eV. Such anti-crossing may be considered as a 2D gapped Dirac point. From symmetry analysis, we construct the following $k\cdot p$ effective model for this gapped Dirac point:
\begin{equation}
  H_\text{eff}=w k_y+v_x k_x\tau_y+v_y k_y\tau_z+m\tau_x,
\end{equation}
where momentum $k$ is measured from the Dirac point, $\tau$'s are Pauli matrices, $w$, $v_i$, and $m$ are real model parameters. It is known that Berry curvature is concentrated around such points, where the interband coherence is strong. This explains the feature that we observe in Fig.~\ref{fig4}(c).

The magnetic anisotropy in 2D CrPS$_{4}$ is not very large.  From our calculations, the energy difference between out-of-plane and in-plane magnetic configurations is about 0.04 meV per Cr atom. It was shown in experiment that the magnetic moment direction can be rotated by relatively small applied magnetic field $\sim 90$ mT~\cite{son2021airb}. Hence, we also consider the AHE response in monolayer when the magnetization is rotated to an in-plane direction. In Fig.~\ref{fig4}(d), we show the calculated $\sigma_{xy}$ for magnetization along the $x$ direction. Again, several peaks can be observed in the figure. The peak values are comparable to those in Fig.~\ref{fig4}(a).
The first peak in valence band has a value $\sim 370$ S/cm. Furthermore, for magnetization along $y$ direction, $\sigma_{xy}$ is identically zero. This is because in this configuration, the vertical mirror $\mathcal{M}_{y}$ symmetry is preserved, which forbids the linear AHE.

\section{nonlinear Hall response in bilayer}
For the even-layer CrPS$_{4}$, the state above N\'{e}el temperature $T_N$ possesses both $\mathcal{P}$ and $\mathcal{T}$ symmetries, which prohibit linear AHE and any second order current response. Below $T_N$, the AFM state preserves $\mathcal{PT}$ symmetry, so that the linear AHE is still suppressed, however, the nonlinear Hall response starts to appear.
This nonlinear response can be expressed as
\begin{equation}
  j_{a}=\chi_{abc}E_{b}E_{c},
\end{equation}
where $\chi_{abc}$ is the nonlinear conductivity tensor, the subscripts label the Cartesian components, and the Einstein summation convention is assumed here. Here, we focus on the intrinsic contribution to nonlinear Hall transport, which is determined solely by the band structure property. It should be noted that the extrinsic contribution from Berry curvature dipole is forbidden by the $\mathcal{PT}$ symmetry here~\citep{PhysRevLett.115.216806}.

\begin{figure}
  \includegraphics[width=8.6cm]{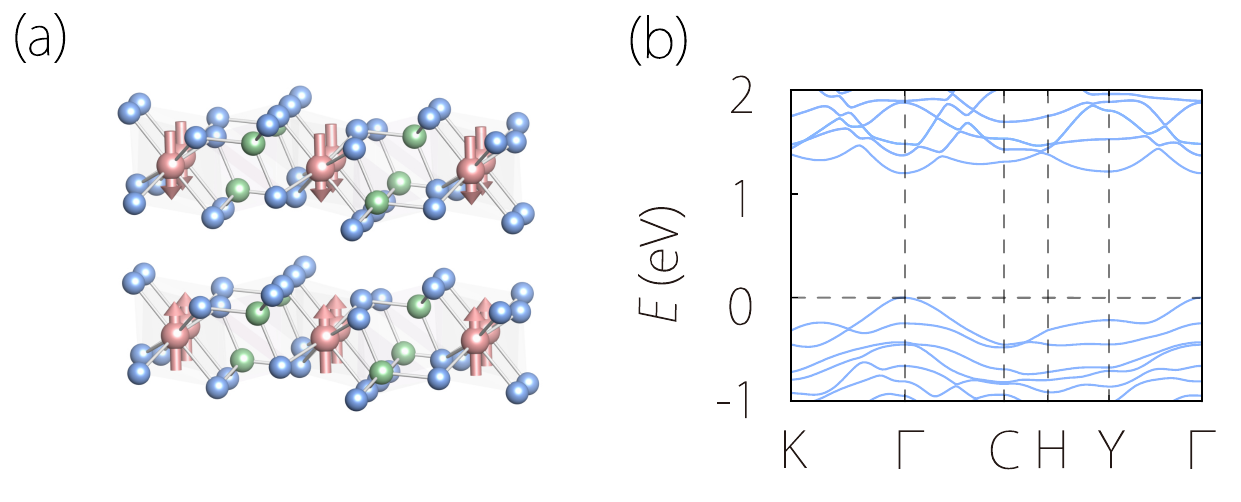}
  \caption{\label{fig5}(a) Lattice structure of bilayer CrPS$_4$, which has $A$-type AFM ground state. (b) Calculated band structure for bilayer CrPS$_4$ in AFM ground state. }
\end{figure}

The theory of intrinsic nonlinear Hall effect has been developed in Ref~\citep{PhysRevLett.127.277201, PhysRevLett.112.166601,PhysRevLett.127.277202}. The corresponding nonlinear conductivity tensor can be expressed as

\begin{equation}\label{chi}
\chi_{abc}=\frac{e^2}{\hbar}\int[d\boldsymbol{k}]f_{n\bm k}\Lambda^n_{abc}(\bm k)
\end{equation}
where
\begin{equation}\label{Lambd}
  \Lambda^n_{abc}(\bm k)=\partial_b G_{ac}^n-\partial_a G_{bc}^n
\end{equation}
is the $k$-resolved BCP dipole, and
\begin{equation}\label{Gtensor}
G_{ab}^{n}(\boldsymbol{k})=2e\hbar^2\mathrm{Re}\sum_{m\neq n}\frac{(v_a)_{nm}(v_b)_{mn}}{(\varepsilon_{n}-\varepsilon_{m})^3}
\end{equation}
is the BCP tensor, and $\partial_a\equiv \partial_{k_a}$. Like Berry curvature, BCP is a gauge-invariant quantity. Physically, it characterizes the shift of a electron wavepacket enter by an applied electric field~\cite{PhysRevLett.112.166601}. Near hot spots formed by a pair of bands, BCP can be further connected to the quantum metric of band structure~\cite{FENG2025100040}. From Eqs.~(\ref{chi}, \ref{Lambd}, \ref{Gtensor}), one can clearly see that the intrinsic response is purely Hall and dissipationless, i.e., $\chi_{abc}$ is antisymmetric in its first two indices~\cite{xiao2025proper}. As a consequence, the intrinsic conductivity $\chi_{abc}$ is also forbidden by any rotational symmetry normal to the transport plane. These key features and the theory have been successfully verified in experiments on MnBi$_2$Te$_4$~\cite{Gao_2023} and Mn$_3$Sn~\cite{han_room-temperature_2024}. Meanwhile, responses from extrinsic mechanisms do not share such characters~\cite{PhysRevB.111.155127, gong2025nonlinear, Wang_2023}. This theory has found successful applications in explaining key features in experiments on MnBi$_2$Te$_4$~\cite{Gao_2023} and Mn$_3$Sn~\cite{han_room-temperature_2024}.

\begin{figure}
  \includegraphics[width=8.6cm]{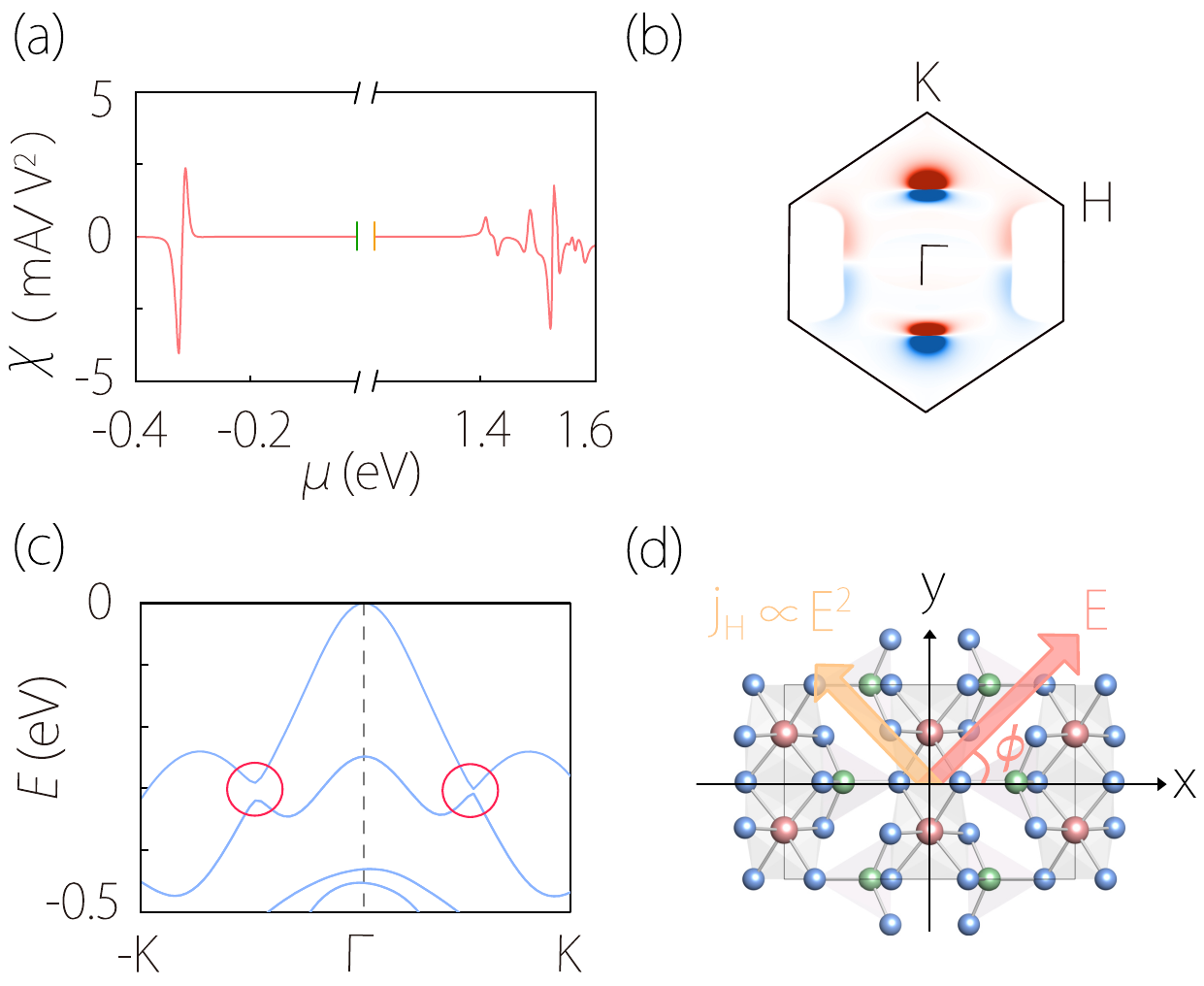}
  \caption{\label{fig6} (a) Calculated intrinsic nonlinear Hall conductivity $\chi_{yxx}$ as a function of chemical potential for bilayer AFM CrPS$_4$. (b) $k$-space distribution of BCP dipole for states below $\mu= -0.3$~eV. The hot spots on $\Gamma$-$K$ path correspond to the two gapped Dirac points marked in the enlarged band structure in (c).(d) Sketch of nonlinear Hall current induced by $E$ field.}
\end{figure}

In the following, we focus on the bilayer CrPS$_{4}$ as a representative. The calculated band structure is plotted in Fig.~\ref{fig5}(b).
It should be noted here that each band is doubly spin degenerate due to the $\mathcal{PT}$ symmetry.
As for the nonlinear conductivity tensor, we note that the $2/m^{\prime}$ symmetry of the system dictates that the intrinsic response is characterized by a single independent component, with
\begin{equation}
  \chi_{yxx}=-\chi_{xyx}.
\end{equation}
For an in-plane $E$ field in the direction $\bm E=E(\hat{x}\cos{\phi}+\hat{y}\sin{\phi})$, where $\phi$ is the polar angle measured from $x$ axis(as shown in Fig.~\ref{fig6}(d)), the resulting Hall current  $j_\text{H}$ in the direction normal to $E$ field can be expressed as
\begin{equation}\label{angle}
  j_\text{H}=\chi_\text{H}(\phi)E^2, \qquad \chi_\text{H}(\phi)=\chi_{yxx}\cos\phi.
\end{equation}

In nonlinear Hall experiment, one typically fabricate a disk shaped device and multiple leads, such that the driving current can be applied along various in-plane directions while measuring the Hall response in the transverse direction~\citep{kang2019nonlinear,lai2021third}. Equation (\ref{angle}) tells us the angular dependence of the measured Hall signal should exhibit a simple cosine dependence, with a $2\pi$ periodicity.

Next, we compute the nonlinear conductivity component $\chi_{yxx}$ from our DFT result. The result is shown in Fig.~\ref{fig6}(a).
One observes that the value on the electron-doping side can reach $\sim0.4$~mA/V$^{2}$. For hole doping, the peak at $\mu=-0.3$~eV is even larger, reaching $\sim2$~mA/V$^{2}$, which is quite large. It is comparable to the one measured in few-layer MnBi$_{2}$Te$_{4}$~\citep{ Gao_2023,gao2024antiferromagnetic}.

To reveal the origin of this peak, we plot the distribution of $k$-resolved BCP dipole $\Lambda_{yxx}(\bm k)$ in BZ by summing over all occupied states up to $\mu=-0.3$ eV. The result is plotted in Fig.~\ref{fig6}(b). One can see a pronounced dipole like distribution in a small region around a point on $\Gamma$-$K$ ($k_y>0$) path. This corresponds to a band anti-crossing point as indicated in Fig.~\ref{fig6}(c). By analyzing the band structure around this point, we identify it as a gapped Dirac point. However, different from that in Fig.~\ref{fig4}(d), the bands are doubly degenerate. The effective model for this point from symmetry analysis takes the following form:
\begin{equation}\label{GDP}\begin{split}
  H_\text{eff}=wk_y +v_x k_x (\sigma_x\cos\alpha +\sigma_y\sin\alpha )\tau_y\\
  +v_y k_y\sigma_z\tau_y+m\tau_x,
  \end{split}
\end{equation}
where the wavevector $k$ is measured from the point, $\sigma$ and $\tau$ are two sets of Pauli matrices, $\tau^{\prime}s$ denote the two doubly degenerate anti-crossing bands, and $\sigma^{\prime}s$ denote the spin degree of freedom within each doubly degenerate band, $w$, $v$'s, $\alpha$, and $m$ are real model parameters. The last term in (\ref{GDP}) opens a small gap $\sim0.007$~eV between the two doubly degenerate bands. As shown in Ref.~\cite{PhysRevLett.127.277201, PhysRevLett.127.277202}, such gapped Dirac point tends to enhance BCP hence makes significant contribution to the nonlinear Hall response.
In Fig.~\ref{fig6}(c), there is another hot spot on the $k_y<0$ portion of $\Gamma$-$K$ path, with a comparatively weaker BCP dipole distribution. This corresponds to another gapped Dirac point at energy slightly below $-0.3$~eV (see Fig.~\ref{fig6}(b)). The form of its effective model is the same as Eq.~(\ref{GDP}), since the local symmetry is the same.

 The above discussion is for the AFM bilayer with top layer spin-down and bottom layer spin-up. As for the other energy-degenerate configuration with top layer spin-up and bottom layer spin-down, the nonlinear Hall response will flip sign. This is because the two magnetic configurations are connected by time reversal operation, and the BCP dipole induced
nonlinear Hall response is a $\mathcal{T}$-odd effect.

\section{Layer Hall effect in bilayer}

We have shown that a large AHE can be realized in monolayer CrPS$_{4}$. The bilayer is made of two monolayers with opposite magnetization. Hence, the vanishing of AHE in bilayer can be intuitively understood as the exact cancellation of the AHE responses of the two monolayers. From this perspective, one may consider the bilayer system actually hosts a layer Hall effect~\cite{gao2021layer}, i.e., the AHE response  is finite in each layer channel and is opposite between the two channels. This is analogous to concepts of the spin Hall effect and valley Hall effect, with layer index being the binary degree of freedom.

Similar to spin or valley Berry curvature, we may define a layer Berry curvature~\citep{fan2024intrinsic} for bilayer CrPS$_{4}$:
\begin{equation}\label{LBC}
  \Omega_{z}^{n,\ell}(\boldsymbol{k})=-2\hbar^2 \text{Im}\sum_{m\neq n} \frac{(v_x^\ell)_{nm}(v_y)_{mn}}{(\varepsilon_n-\varepsilon_m)^2},
\end{equation}
where $\ell=T,B$ is the binary layer index for top and bottom layers, $(v_x^\ell)_{nm}$ is the interband matrix element for the operator $\hat v_x^\ell=\{\hat P_\ell,\hat v_x\}/2$, i.e., the velocity operator combined with the projector $\hat P_\ell$ into layer $\ell$. The response of AHE in each layer is characterized by
\begin{equation}\label{LAHE}
\sigma_{xy}^\ell=\frac{e^2}{\hbar}\int[d\boldsymbol{k}]f_{n\bm k}\Omega_{z}^{n,\ell}(\boldsymbol{k}).
\end{equation}
One can easily see that the total anomalous Hall conductivity
\begin{equation}
  \sigma_{xy}=\sigma_{xy}^T+\sigma_{xy}^B,
\end{equation}
must vanish. However, the layer Hall conductivity
\begin{equation}
    \sigma_{xy}^L=\sigma_{xy}^T-\sigma_{xy}^B
\end{equation}
can be nonzero.

\begin{figure}[t!]
  \includegraphics[width=8.6cm]{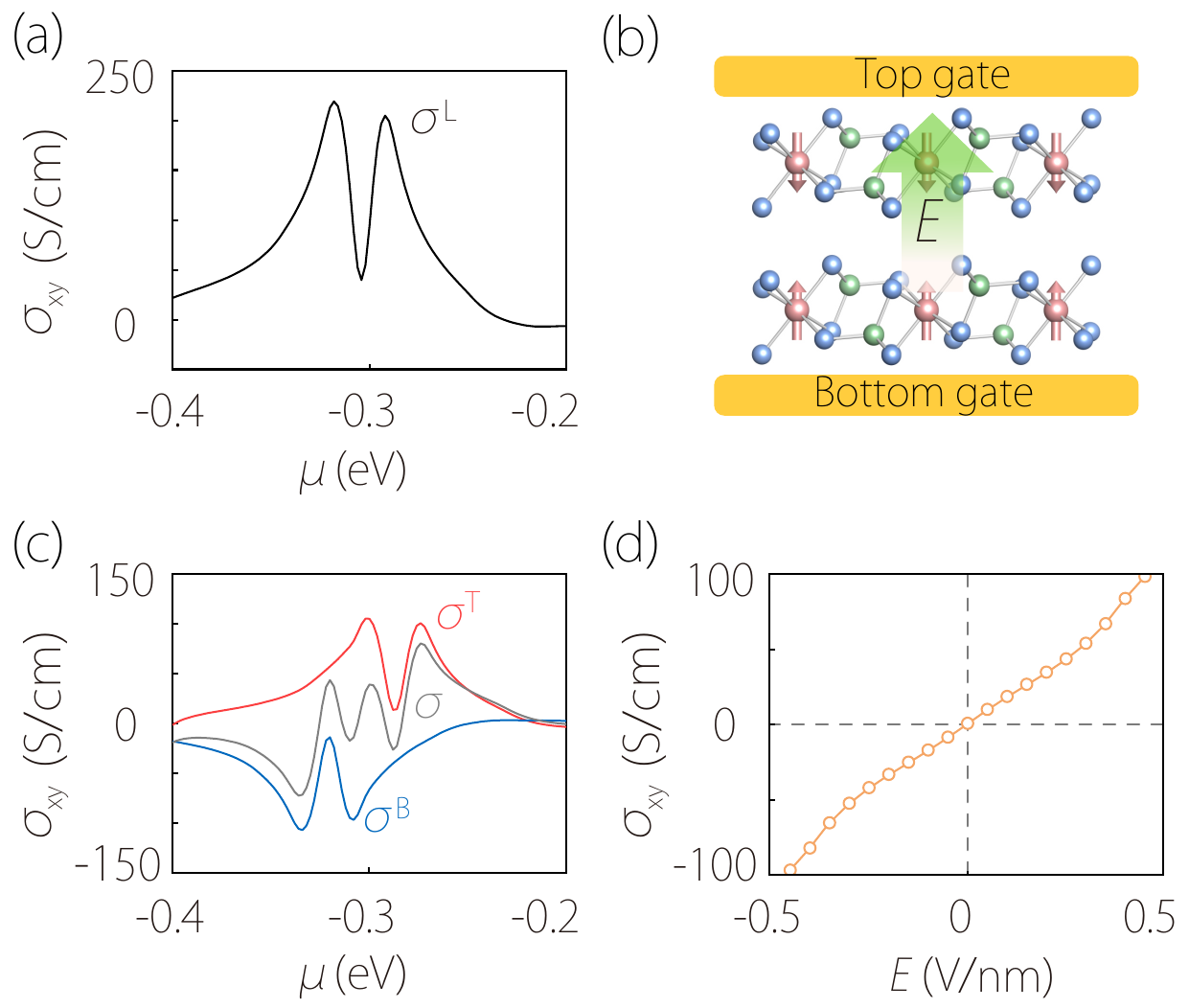}
  \caption{\label{fig7} (a) Calculated layer Hall conductivity in bilayer CrPS$_4$. Here, we focus on the peak structure in valence band. (b) Schematic figure showing the application of a gate electric field to  bilayer CrPS$_4$. This breaks the $\mathcal{PT}$ symmetry and generates a finite  anomalous Hall response. {(c) Layer-resolved Hall conductivities and total Hall conductivities under a gate field of $0.02$~V/nm.} (d) The resulting  anomalous Hall conductivity versus gate field at a fixed chemical potential of $-0.25$~eV.}
\end{figure}

In Fig.~\ref{fig7}(a), we show the calculated layer Hall conductivity for bilayer CrPS$_{4}$. One can see its peak value can reach $\sim 220$~S/cm
on the hole doping side. This indicates that although the net charge current flow vanishes, the AHE response within each layer is still quite large, leading to a large layer Hall effect.

It is noted that the existence of layer Hall effect does not require the breaking of $\mathcal{PT}$ symmetry. Nevertheless, direct detection of layer resolved Hall response is difficult. To probe layer Hall effect in experiment, an indirect way is to break the $\mathcal{PT}$ symmetry that connects the two layers by gating, so that the layer Hall effect can be converted to an anomalous Hall signal~\cite{gao2021layer}.Figure \ref{fig7}(c) shows the calculated layered-resolved and net AHE conductivities for bilayer CrPS$_{4}$ under a gate field of $ 0.02$~V/nm (a moderate gate field strength that can be readily achieved in experiment). One observes that the gate field shifts the layered-resolved conductivities $\sigma_{xy}^\ell$ in different ways, leading to a finite AHE. In Fig.~\ref{fig7}(d), we plot the net AHE conductivity as a function of gate field at a fix chemical potential $\mu=-0.25$~eV. One observes the linear scaling at  weak field region, and the sign of AHE response can be controlled by the direction of gate field, which is a nice feature for device applications.

\section{In-plane anomalous Hall effect in bilayer}

As mentioned above, even-layer CrPS$_{4}$ should have vanishing linear AHE due to the $\mathcal{PT}$ symmetry.
By applying an in-plane magnetic field, the $\mathcal{PT}$ symmetry can be broken, and a linear charge Hall signal can be produced~\cite{cao2023plane, PhysRevB.108.L121301,PhysRevResearch.5.023138,PhysRevB.109.155408}. (Out-of-plane $B$ field also works, but it will bring in the ordinary Hall effect.)
In literature, such a configuration is commonly called the planar Hall effect. However, it should be noted that in many previously reported experiments, the transverse signal is even in the applied magnetic field $B$, which is actually not a Hall response but a transverse magneto-resistance.  This is in contrast to the effect considered here, which is odd in $B$. Because of this difference, some recent experimental studies termed this effect as in-plane AHE~\cite{zhou_heterodimensional_2022,PhysRevLett.132.106601,liu_crystal_2024}, which we adopt here.

The main effect of applied in-plane $B$ field here is to tilt the local magnet moments and cause a net in-plane magnetization $\delta M$ along the $B$ field direction. This picture was confirmed in experiment, and it was found that the magnetic anisotropy in  CrPS$_{4}$ is relatively weak, so the spin canting can be readily achieved by moderate $B$ field strength~\cite{peng2020magneticbbbb}.
The direction of in-plane $B$ field affects the induced AHE. For field along $y$ direction, the resulting magnetization $\delta M_y$ cannot efficiently generate AHE, because it preserves the $\mathcal{M}_y$ symmetry for each layer. In comparison, the component $\delta M_x$ along $x$ direction is more effective in producing the in-plane AHE response. In the following, we shall focus on this case, with $B$ field causing spin canting in the $x$ direction.

\begin{figure}
  \includegraphics[width=8.6cm]{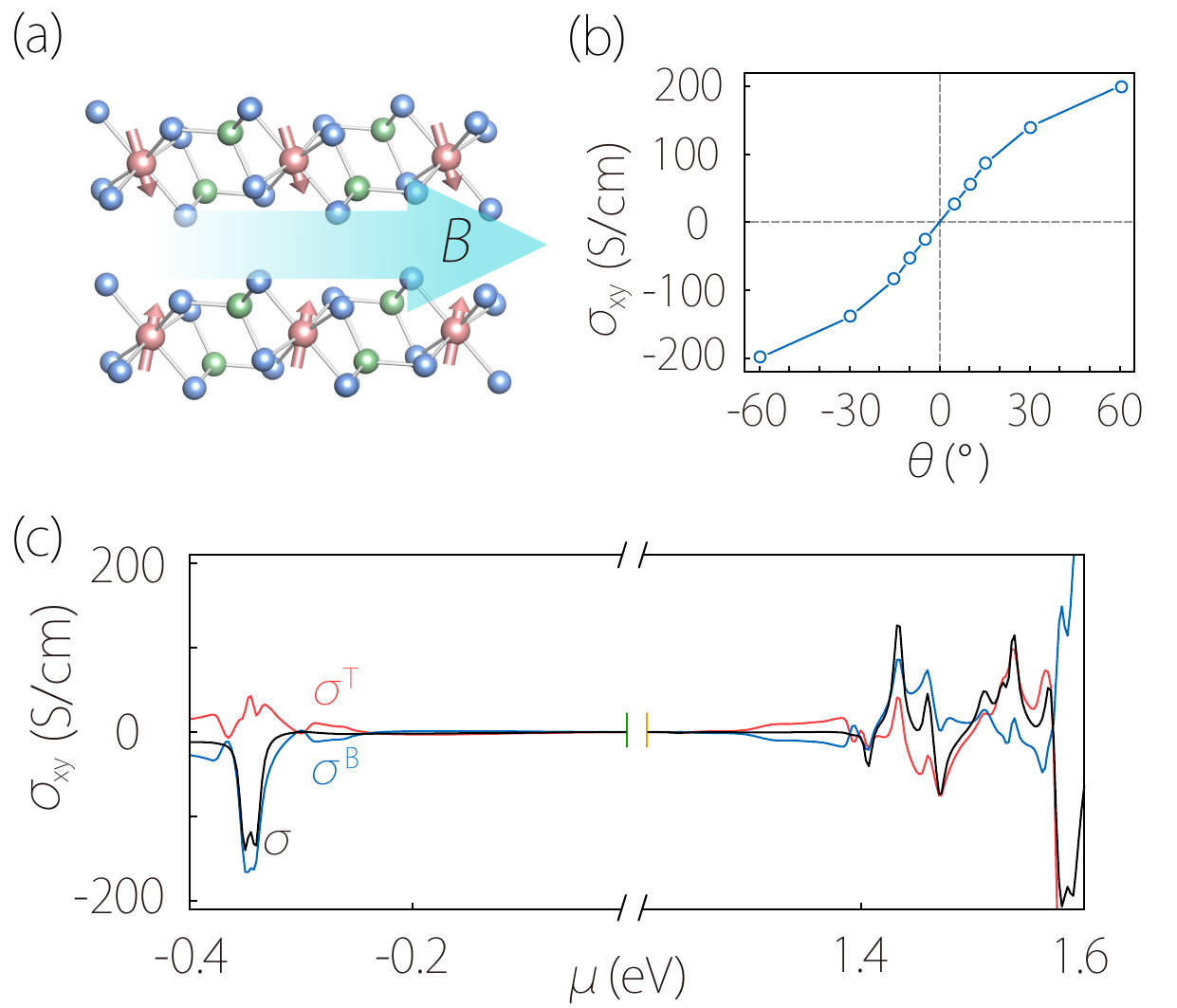}
  \caption{\label{fig8}(a) Schematic figure showing the application of an in-plane magnetic field to bilayer CrPS$_4$.
  The field breaks $\mathcal{PT}$ symmetry, tilts the local moments, and induces a finite linear anomalous Hall response.
  (b) Calculated anomalous Hall conductivity versus spin-canting angle $\theta$ at a fixed chemical potential of $-0.31$~eV. (c) Calculated layer-resolved and total anomalous Hall conductivities for the bilayer when the spin-canting angle $\theta = 30 ^{\circ}$.  }
\end{figure}

In DFT calculation, we model the spin-canted state by rotating the local moments towards the $x$ direction by a canting angle $\theta$. At a small angle $\theta= 30^{\circ}$, the calculated layered-resolved and net AHE conductivities are plotted in Fig.~\ref{fig8}(c). One observes that the contributions from the two layers no longer cancel out. A sizable AHE conductivity with peak value $\sim 132$~S/cm can be obtained. In Fig.~\ref{fig8}(b), we plot the induced AHE conductivity as a function of canting angle  $\theta$
for a fixed $\mu= -0.31$~eV. The curve shows a linear increase at small angles and tends to saturate at large angle. It reaches $\sim 201$~S/cm at angle $\theta= 60^{\circ}$.
 Using a simple AFM model, the magnetic energy under an in-plane magnetic field can be written as
$E_m =-2g\mu_B BS\sin\theta -J_c S^2 \cos 2\theta$, where $S$ is the magnitude of local spin, and $J_c$ is the interlayer AFM coupling strength~\cite{Daniel2010}. In this estimation, the magnetic anisotropy energy is neglected, since it is two order of magnitude smaller than the exchange energy~\cite{son2021airb}.
Minimizing this energy with respect to $\theta$, we find the magnitude of $B$ field needed to achieve  canting angle $\theta$ can be estimated as 
\begin{equation}
  B\approx 2J_{c}S\sin(\theta)/(g\mu_{\mathrm{B}}).
\end{equation} 
Taking the interlayer exchange coupling $J_{c}=0.15$~meV as obtained from experiment~\cite{son2021airb}, and $S=3/2$, we find that a canting angle of $5^{\circ}$ can be achieved by a moderate $B$ field of $340$~mT, showing a pronounced in-plane AHE response in bilayer CrPS$_{4}$.

\section{Discussion and Conclusion}

In this study, we have investigated the rich anomalous Hall transport phenomena in 2D CrPS$_{4}$ systems.
Monolayer and bilayer CrPS$_{4}$ are used as representatives for demonstration. The general features apply also to
other few-layer CrPS$_{4}$ systems. The odd layers share the same symmetry as monolayer, while the even layers share the same symmetry as bilayer. Hence, linear AHE is expected in odd layers, whereas intrinsic nonlinear Hall effect is expected in even layers.
The linear AHE induced by gate field or in-plane magnetic field should occur in even-layer CrPS$_{4}$ as well.
It is worth noting that the nonlinear Hall response would vanish in the bulk limit, since bulk CrPS$_{4}$
retains inversion symmetry regardless of the direction of its N\'{e}el vector.

Experimentally, 2D CrPS$_{4}$ has been successfully fabricated via various techniques down to monolayer limit~\cite{son2021airb,Lee2017Structural}. Field effect transistor devices based on 2D CrPS$_{4}$ have also been demonstrated~\cite{wu2022magnetotransport,wu2023gate, wu2023magnetism,cheng2024quantum}. The phenomena predicted here should be detectable via standard transport measurement with a dual gate device setup. For the nonlinear Hall effect, the typical method is to modulate the driving current with a low frequency (less than 100 Hz) and measure the response signal at second harmonic frequency by using the lock-in technique~\cite{ma_observation_2019,kang2019nonlinear}. The angular dependence of the signal (as manifested in Eq.~(\ref{angle})) can be probed by fabricating a disk shaped device with multiple leads to allow driving current (and response signal) exerted (detected) along different in-plane directions. As for the layer Hall effect in bilayer, besides detection via charge Hall signal under electric gating, it induces an in-plane orbital magnetization~\cite{fan2024intrinsic} and hence
may also be probed by magneto-optical Kerr microscopy~\cite{huang2017layer, gong2017discovery}.

For odd-layer CrPS$_{4}$, the nonlinear transport signal
should vanish both above and below magnetic transition temperature $T_c$. For even layers,
the nonlinear signal is forbidden above $T_c$ (since the crystal structure preserves $\mathcal{P}$) and appears  only when temperature is lowered below $T_c$. This could be used as a signature for detecting $T_c$ or the parity of number of layers.
In addition, we note that for temperature above $T_c$, both linear and second-order Hall effects are forbidden by the $\mathcal{T}$ and $\mathcal{P}$ symmetries. The leading order anomalous Hall response should be from the third order  ~\cite{PhysRevB.105.045118,lai2021third,wang2022room,PhysRevB.106.045414,PhysRevB.106.035307,PhysRevB.107.205120,PhysRevB.107.245141,PhysRevLett.131.186302}, which was recently probed in several nonmagnetic materials.

In this work, we focus on the intrinsic responses, which represent inherent material properties. There also exist extrinsic contributions arising from disorder scattering, which vary from sample to sample~\cite{PhysRevLett.131.076601,PhysRevB.108.L201115,PhysRevB.111.155127}. Experimentally, a typical method to distinguish the different contributions is to perform a scaling analysis by varying the various disorder strengths.
The intrinsic response correspond to a component that is independent of scattering. And first-principles result serves as an important benchmark for such analysis. As we have mentioned, even-layer CrPS$_{4}$ represents a good platform for studying intrinsic nonlinear AHE, because of the following reasons: (1) It low-symmetry monoclinic structure makes the
intrinsic nonlinear AHE a property of the magmatic material itself, without requiring interfacial effects as in previous experiments; (2) Its spacetime inversion symmetry  suppresses the extrinsic contributions, making the intrinsic response more important; and (3) its 2D vdW character makes the system highly tunable and readily integrated into vdW heterostructures for nonlinear device applications.

In conclusion, we have investigated various anomalous Hall transport phenomena in 2D CrPS$_{4}$.
Due to their low crystal symmetry and layer-dependent magnetism, we find a rich variety of anomalous Hall responses, with sizable conductivities predicted by our first-principles calculations. These effects allow us to probe the intrinsic band geometric quantities, such as Berry curvature, BCP, and quantum metric, of the material system.
The demonstrated controllability of Hall transport in
2D CrPS$_{4}$ also offers possible mechanisms for device operations.
Our predications here can be readily tested in experiment, and they may find promising applications in next-generation electronics and spintronics based on 2D magnetic materials.

\begin{acknowledgments}
  The authors thank D. L. Deng for valuable discussions. C.X. acknowledges the support by the start-up funding from Fudan University. This work is supported by UM Start-up Grant (SRG2023-00033-IAPME, SRG2022-00030-IAPME, and SRG2023-00057-IAPME), UM Multi-Year Research Grant (MYRG-GRG2023-00206-IAPME-UMDF), Science and Technology Development Fund of Macau SAR (0048/2023/RIB2 and 0066/2024/RIA1), National Key R$\&$D Program of China (Grant No.~2022YFA1402600), NSFC (Grants No.~12174018) and HK PolyU Start-up Grant (P0057929).
\end{acknowledgments}

\bibliographystyle{apsrev4-2}
\bibliography{ref.bib}

\end{document}